\documentclass{elsart}
\usepackage{amssymb}
\usepackage{amsmath}
\usepackage{bm}
\usepackage{graphicx}
\usepackage{psfrag}
\usepackage{aas_macros}

\newcommand{\vecb}[1]{{\bm{\mathrm{#1}}}}
\newcommand{\Dfrac}[2]{\frac{\mathrm{d}{#1}}{\mathrm{d}{#2}}}

\begin{document}

\begin{frontmatter}

\title{An efficient method for calculation of cooling in Lagrange computational gas
  dynamics}

\author{E.~P. Kurbatov}
\ead{kurbatov@inasan.ru}
\address{Institute of Astronomy, Moscow, Russia}

\begin{abstract}
A new method for computation of gas cooling for Lagrange approach is
suggested.  The method is based on precalculation of cooling law for known cooling
function.  Unlike implicit methods, this method is very efficient, it is an
one-step method which is even more accurate than implicit methods of the same
order.
\end{abstract}

\begin{keyword}
gas dynamics \sep cooling

{\it ZMATH:} 65L05, 85-08, 85A30

\PACS 02.60.Cb \sep 47.11.-j
\end{keyword}

\end{frontmatter}

\section{Introduction}
\label{sec:introduction}

The Lagrange approach in gas dynamics assumes solution of the energy conservation
equation in the form
\begin{equation}
  \Dfrac{u}{t} = -\frac{P}{\rho}\,\nabla\vecb{v} - \frac{\Lambda}{\rho}  \;,
  \label{eq:energy_conservation_law}
\end{equation}
where $u$ is the internal energy per unit mass, $P$ is the pressure,
$\vecb{v}$ is the gas velocity, $\Lambda$ is the cooling rate per unit volume
that depends on gas density $\rho$ and  temperature $T$.  To solve
(\ref{eq:energy_conservation_law}) the implicit methods are usually used,
because the characteristic cooling time can be much smaller than gas-dynamical
time.  This circumstance makes the usage of explicit methods impossible.

An efficient one-step method for cooling computation for the energy
conservation law in the form (\ref{eq:energy_conservation_law}) is suggested in this
paper.  It is implied for the use in the computational gas dynamics and it is intended to
predict variation of the gas temperature over a  time step.  This method utilizes the
information on given cooling function $\Lambda(\rho, T)$ and uses approximation
in which cooling time is shorter than gas-dynamical time.  The last condition
is fulfilled in many gas-dynamical problems, particularly, in astrophysical problems which
are in the area of the author's interest.

\section{One-step method}
\label{sec:method}

Since cooling function depends on the temperature, let rewrite the equation
(\ref{eq:energy_conservation_law}) in the temperature and density terms.  Assuming
the equation of state of an ideal gas with mean molecular weight $\mu$ and
adiabatic exponent $\gamma$, equation (\ref{eq:energy_conservation_law}) gives
\begin{equation}
  \Dfrac{T}{t} = -(\gamma-1) \left[ (\nabla\vecb{v})\,T
    + \frac{\mu}{k_\mathrm{B}}\,\frac{\Lambda}{\rho} \right]  \;,
  \label{eq:temperature-eq}
\end{equation}
where $k_\mathrm{B}$ is the Boltzmann constant.  The first term in the
right-hand side of this equation conforms to purely adiabatic temperature
change while the second one corresponds to cooling process.  Following 
Sutherland \& Dopita \cite{Sutherland--1993ApJS...88..253S} and Nakasato et
al. \cite{Nakasato--2000ApJ...535..776N} we suppose cooling function to be
proportional to the squared number density $\Lambda = n^2 \Lambda^\ast$, while
$\Lambda^\ast$ depends on the logarithmic temperature only:
\begin{equation}
  \Dfrac{\log T}{t} = -\frac{\gamma-1}{\ln 10} \left[ \nabla\vecb{v}
    + \frac{\rho}{\mu k_\mathrm{B}}\,\frac{\Lambda^\ast(\log T)}{T} \right]
    \;,
  \label{eq:log-temperature-eq}
\end{equation}
where the definition $\rho = \mu n$ was used.

Despite of condition imposed on characteristic times,  adiabatic cooling or
heating may  contribute significantly to temperature change.  To account for
this possibility and, simultaneously,  to simplify the solving procedure, let to
split the right-hand side of equation (\ref{eq:log-temperature-eq}) into two
independent processes, adiabatic and cooling ones.  Since cooling time is
short compared to gas-dynamical time, the velocity divergence and density may be
taken as a constants.  The resulting approximate expression for logarithmic
temperature variation is
\begin{equation}
  \log T = -\frac{\gamma-1}{\ln 10}\,(\nabla\vecb{v})\,t + \sigma  \;,
  \label{eq:log-temperature}
\end{equation}
where $\sigma$ satisfies the equation
\begin{equation}
  \Dfrac{\sigma}{t} = -\frac{\gamma-1}{\ln 10}\,
    \frac{\rho}{\mu k_\mathrm{B}}\,10^{-\sigma} \Lambda^\ast(\sigma)
  \label{eq:sigma-eq}
\end{equation}
with the initial condition $\sigma(t = 0) = \sigma_0 \equiv \log T_0$, where
$T_0$ is the initial temperature.  Solution of this equation can be written
implicitly in the form
\begin{equation}
  \int_{\sigma_0}^\sigma \frac{\mathrm{d}x}{10^{-x} \Lambda^\ast(x)}
    = -\frac{\gamma-1}{\ln 10}\,\frac{\rho}{\mu k_\mathrm{B}}\,t  \;.
  \label{eq:sigma}
\end{equation}

Equation (\ref{eq:sigma-eq}) is an autonomous one.  It means that the
variation of $\sigma$ depends on variation of $t$ only.  It's possible to use
this property of equation (\ref{eq:sigma-eq}) to precalculate the solution of
(\ref{eq:sigma-eq}).  First, assume the cooling function $\Lambda^\ast$ to be
defined in the  temperature interval $[T_\mathrm{min}, T_\mathrm{max}]$ or, in our
case, in the corresponding interval $[\sigma_\mathrm{min}, \sigma_\mathrm{max}]$.
Second, scale the time variable as
\begin{equation}
  \tau = \frac{\gamma-1}{\ln 10}\,\frac{\rho}{\mu k_\mathrm{B}}\,t  \;.
  \label{eq:tau}
\end{equation}
And third, bin the values of integral (\ref{eq:sigma}) with lower bound
fixed to $\sigma_\mathrm{min}$ and upper bound running from
$\sigma_\mathrm{min}$ to $\sigma_\mathrm{max}$:
\begin{equation}
  \int_{\sigma_\mathrm{min}}^{\sigma_i}
    \frac{\mathrm{d}x}{10^{-x} \Lambda^\ast(x)}
    = -\tau_i  \;.
  \label{eq:binned-sigma}
\end{equation}
This binned function will represent the cooling process over entire range of
temperatures where cooling function is defined.

\begin{figure}
  \begin{center}
    \psfrag{lg T_a}{\scriptsize $\hspace{3mm}\sigma_0$}
    \psfrag{lg T_b}{\scriptsize $\hspace{3mm}\sigma$}
    \psfrag{lg t_a}{\scriptsize $\log t_\mathrm{a}$}
    \psfrag{lg t_b}{\scriptsize $\log t_\mathrm{b}$}
    \includegraphics*[width=9cm]{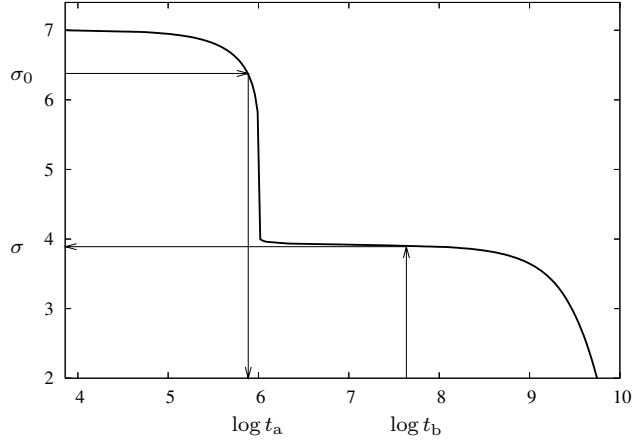}
  \end{center}
  \caption{The way to get the logarithmic temperature variation for given
  initial logarithmic temperature $\sigma_0$ and time interval $t_\mathrm{b} -
  t_\mathrm{a}$.  The predicted value of $\sigma$ used in
  (\ref{eq:log-temperature}) for temperature prediction.}
  \label{fig:sigma-t}
\end{figure}

An example of how to find $\sigma$ for given initial logarithmic temperature
$\sigma_0$ and time interval is shown in fig. (\ref{fig:sigma-t}).
Here the cooling function of Sutherland \& Dopita
\cite{Sutherland--1993ApJS...88..253S} (for $T > 10^4$ K) and Nakasato et
al. \cite{Nakasato--2000ApJ...535..776N} (for $T < 10^4$ K) was used (the
function plotted corresponds to heavy elements abundance  five
orders of magnitude lower than the solar one).  For initial logarithmic
temperature $\sigma_0$ the corresponding initial time $t_\mathrm{a}$ in binned
function (\ref{eq:binned-sigma}) is searched first.  Next, using the binned
inverse function, the final value of $\sigma$ is obtained for time moment,
shifted to prediction time $t_\mathrm{b} = t_\mathrm{a} + t$.  This algorithm
needs two arrays, for logarithmic temperature and for scaled time.  The actual
values of both functions can be obtained by interpolation procedure defined
for  these arrays, i.e. just two array lookups are needed to get the temperature value
by  the expression (\ref{eq:log-temperature}).

It is interesting to estimate the error of the one-step method compared to the implicit
method for solution of (\ref{eq:energy_conservation_law}) or
(\ref{eq:temperature-eq}).  Let denote $T_\mathrm{os}$ the solution derived by
one-step method and $T_\mathrm{im}$ the solution derived by implicit method
\begin{equation}
  \frac{T_\mathrm{im} - T_0}{t}
  = -(\gamma-1) \left[ (\nabla\vecb{v})\,T_\mathrm{im}
    + \frac{\rho}{\mu k_\mathrm{B}}\,\Lambda^\ast(\log T_\mathrm{im}) \right]
  \;.
\end{equation}
It's easy to show that the errors of both methods arise in the second order.  The
second derivatives of deviation of approximate solution from the exact value are:
\begin{equation}
  \epsilon_\mathrm{os} \equiv
  \left| \Dfrac{^2}{t^2}\left(\ln T_\mathrm{os} - \ln T\right) \right|_{T_0}
  = \frac{1}{\tau_\mathrm{gd}\,\tau_\mathrm{cool}}\,
    \left. \Dfrac{\log(\Lambda^\ast/T)}{\log T} \right|_{T_0}  \;,
  \label{eq:one-step-method-error}
\end{equation}
\begin{multline}
  \epsilon_\mathrm{im} \equiv
  \left| \Dfrac{^2}{t^2}\left(\ln T_\mathrm{im} - \ln T\right) \right|_{T_0} \\
  = \left( \frac{1}{\tau_\mathrm{gd}} + \frac{1}{\tau_\mathrm{cool}} \right)^2
    + \frac{1}{\tau_\mathrm{cool}^2}\,
        \left. \Dfrac{\log(\Lambda^\ast/T)}{\log T} \right|_{T_0}
    + \epsilon_\mathrm{os}  \;,
  \label{eq:implicit-method-error}
\end{multline}
where the gas-dynamical time and cooling time are introduced:
\begin{equation}
  \frac{1}{\tau_\mathrm{gd}} = (\gamma-1)\,|\nabla\vecb{v}|  \;,\qquad
  \frac{1}{\tau_\mathrm{cool}} = (\gamma-1)\,\frac{\rho}{\mu k_\mathrm{B}}\,
    \frac{\Lambda^\ast(\log T_0)}{T_0}  \;.
\end{equation}
It is clearly seen that the error of implicit method is systematically larger while the
one-step method gives exact solution in pure adiabatic or pure cooling limit.

\section{Conclusions}

The presented method of calculation of cooling in Lagrange computational gas
dynamics is faster and more precise than the implicit method of the same order.

\section{Acknowledgements}

Thanks to Peter Berczik for providing tabulated cooling function and to Lev
Yungelson for correcting manuscript.

This work was supported Russian Foundation for Basic Research grants
05-02-39005-GFEN\_a and 07-02-00454-a.


\bibliography{paper}
\bibliographystyle{elsart-num}

\end{document}